\documentstyle[12pt]{article}

\begin{document}
\begin{center}
{\Large \bf Polarization of vacuum in the decays of neutral kaons }
\bigskip

{\large D.L.~Khokhlov}
\smallskip

{\it Sumy State University, R.-Korsakov St. 2, \\
Sumy 40007, Ukraine\\
E-mail: khokhlov@cafe.sumy.ua}
\end{center}

\begin{abstract}
It is considered the state of $K^0$ before the decay as that
embedded into the vacuum of $+ \bar K^0 - \bar K^0$.
Since the state of $K^0 _S$ has the lifeitme $\tau_S$, and
the state of $K^0 _L$ has the lifeitme $\tau_L$,
the vacuum of $+ \bar K^0 - \bar K^0$ is polarized.
Due to the polarization of the vacuum,
$2\tau_S/\tau_L$ of $K^0_S$ decay with the delay $\approx \tau_L$.
Thus CP-invariance holds true in the decays of neutral kaons.

\end{abstract}

As known~\cite{Co} strange $K^0$ and $\bar K^0$ mesons do not
possess surtain lifetimes relative to the weak decays, since
weak interactions do not conserve the strangeness number.
There exist two
independent linear combinations of $K^0$ and $\bar K^0$
\begin{equation}
|K^0_S>=\frac{1}{\sqrt 2} (K^0 - \bar K^0)
\label{eq:KS}
\end{equation}
\begin{equation}
|K^0_L>=\frac{1}{\sqrt 2} (K^0 + \bar K^0)
\label{eq:KL}
\end{equation}
They correspond to the particles with the lifetimes
$\tau_S=8.9\times 10^{-11}\ {\rm s} $ and
$\tau_L=5.2\times 10^{-8}\ {\rm s} $ respectively.
The states $K^0_S$ and $K^0_L$ are
of CP-invariance with the eigenvalues +1 and -1 respectively.
$K^0_S$ decays into the system of two pions
\begin{equation}
K^0_S \rightarrow \pi\pi
\label{eq:KSd}
\end{equation}
with the CP-eigenvalue +1,
and $K^0_L$
decays into the system of three pions
\begin{equation}
K^0_L \rightarrow \pi\pi\pi
\label{eq:KLd}
\end{equation}
with the CP-eigenvalue -1.

$K^0$ is considered as a superposition of
$K^0_S$ and $K^0_L$
\begin{equation}
|K^0>=\frac{1}{\sqrt 2}(K^0_S + K^0_L).
\label{eq:K0}
\end{equation}
Evolution of $K^0$ is given by
\begin{equation}
|K^0(t)>=\frac{1}{\sqrt 2}(K^0_S e^{-t/2\tau_S} +
K^0_L e^{-t/2\tau_L}).
\label{eq:K0t}
\end{equation}
One can expect that, within the time $t<\tau_S$, $K^0$ decays
into two pions, and within the time $\tau_S<t<\tau_L$, $K^0$ decays
into three pions. But, within the time $\tau_S<t<\tau_L$,
there exists the probability of the decays
of $K^0$ into two pions
\begin{equation}
\frac{\Gamma(K^0(\tau_S<t<\tau_L)\rightarrow \pi^+\pi^-)}
{\Gamma(K^0(\tau_S<t<\tau_L)\rightarrow all)}
\approx 2\times 10^{-3}
\label{eq:GG}
\end{equation}
\begin{equation}
\frac{\Gamma(K^0(\tau_S<t<\tau_L)\rightarrow \pi^0\pi^0)}
{\Gamma(K^0(\tau_S<t<\tau_L)\rightarrow all)}
\approx 10^{-3}.
\label{eq:GG0}
\end{equation}
Decays $K^0\rightarrow \pi\pi$ within the time $\tau_S<t<\tau_L$
are treated as CP-violation.

Consider the decay of $K^0$. In view of (\ref{eq:KS}),
(\ref{eq:KL}), (\ref{eq:K0}),
$K^0$ decays in combination with $- \bar K^0$ as $K^0 _S$
and in combination with $+ \bar K^0$ as $K^0 _L$. From this
one can consider $K^0$ before the decay as that embedded into
the vacuum of $+ \bar K^0 - \bar K^0$.
Since the probabilities of the decay of $K^0$ in combinations
with $+\bar K^0$ given by $1/\tau_L$ and in combinations with
$-\bar K^0$ given by $1/\tau_S$ are different,
the vacuum of $+ \bar K^0 - \bar K^0$ is polarized.
Due to the polarization of the vacuum
$K^0$ is surrounded by $\tau_S/\tau_L$ of $+ \bar K^0$ and
by $1/2-\tau_S/\tau_L$ of $- \bar K^0$.
The decay of $K^0$ occurs via the birth of the pair
of $+ \bar K^0 - \bar K^0$ from vacuum.
From this the decay of $K^0$ in combination with $+ \bar K^0$
is accompanied by the birth of $- \bar K^0$.
The decay of $K^0$ in combination with $- \bar K^0$
is accompanied by the birth of $+ \bar K^0$.
Taking into account that combination of $K^0$ with $- \bar K^0$
corresponds to the state of $K^0_S$ and combination of $K^0$
with $+ \bar K^0$ corresponds to the state of $K^0_L$,
write the state of $K^0$ in the form
\begin{eqnarray}
|K^0>= & \displaystyle\frac{1}{\sqrt 2} &
\left[\left(1-2\frac{\tau_S}{\tau_L}\right) K^0_S +
\left(2\frac{\tau_S}{\tau_L}\right) K^0_L \right] + \nonumber \\
& \displaystyle\frac{1}{\sqrt 2} &
\left[\left(1-2\frac{\tau_S}{\tau_L}\right) K^0_L +
\left(2\frac{\tau_S}{\tau_L}\right) K^0_S \right].
\label{eq:K01}
\end{eqnarray}
$1-2\tau_S/\tau_L$ of $K^0_S$ and $2\tau_S/\tau_L$ of $K^0_L$
start to decay at $t=0$.
$1-2\tau_S/\tau_L$ of $K^0_L$ start to decay after the decay of
$1-2\tau_S/\tau_L$ of $K^0_S$, i.e. with the delay $\approx \tau_S$.
$2\tau_S/\tau_L$ of $K^0_S$ start to decay after the decay of
$2\tau_S/\tau_L$ of $K^0_L$, i.e. with the delay $\approx \tau_L$.

From the above consideration it follows that, within the time $t<\tau_S$,
$1-2\tau_S/\tau_L$ of $K^0_S$ decay. That is, within the time $t<\tau_S$,
the number of $K^0$ decayed into two pions is estimated as
\begin{equation}
\frac{\Gamma(K^0(t<\tau_S)\rightarrow \pi\pi)}
{\Gamma(K^0(t<\tau_L)\rightarrow all)}=
\frac{1}{2}-\frac{\tau_S}{\tau_L}.
\label{eq:fr2}
\end{equation}
Due to the delay $\approx \tau_L$ caused by the polarization
of the vacuum, within the time $\tau_S<t<\tau_L$,
$2\tau_S/\tau_L$ of $K^0_S$ decay.
That is, within the time $\tau_S<t<\tau_L$,
the number of $K^0$ decayed into two pions is estimated as
\begin{equation}
\frac{\Gamma(K^0(\tau_S<t<\tau_L)\rightarrow \pi\pi)}
{\Gamma(K^0(\tau_S<t<\tau_L)\rightarrow all)}
=2\frac{\tau_S}{\tau_L}=3.4\times 10^{-3}.
\label{eq:G}
\end{equation}

Thus the decays of $K^0$ into two pions
within the time $\tau_S<t<\tau_L$ can be explained by the delay
due to the polarization of the vacuum of $+ \bar K^0 - \bar K^0$.
Hence CP-invariance holds true in the decays of neutral kaons.

\end{document}